\documentclass[aip,jmp,preprint]{revtex4-1}

\usepackage{amsmath}
\usepackage{amsthm}
\usepackage{amssymb}
\usepackage{amscd}
\usepackage{bm}
\usepackage{latexsym}
\usepackage{mathtools}
\usepackage{xcolor}

\usepackage{graphicx}
\usepackage{epsfig}
\usepackage{epstopdf}
\usepackage{subfigure}

\renewcommand{\vec}[1]{\bar{#1}}

\providecommand{\nvec}[1]{\hat{#1}}

\providecommand{\eqref}[1]{(\ref{#1})}

\begin{document}

\preprint{ }

\title{A NOVEL APPROACH TO DESCRIBE CHEMICAL ENVIRONMENTS IN HIGH-DIMENSIONAL NEURAL NETWORK POTENTIALS}

\author{Emir Kocer}
\affiliation{Department of Mechanical Engineering, Bogazici University, Istanbul, TURKEY}
\author{Jeremy K. Mason}
\affiliation{Department of Materials Science and Engineering, University of California Davis, CA, USA}
\author{Hakan Erturk}
\affiliation{Department of Mechanical Engineering, Bogazici University, Istanbul, TURKEY}

\begin{abstract}
A central concern of molecular dynamics simulations are the potential energy surfaces that govern atomic interactions. These hypersurfaces define the potential energy of the system, and have generally been calculated using either pre-defined analytical formulas (classical) or quantum mechanical simulations (ab initio). The former can accurately reproduce only a selection of material properties, whereas the latter is restricted to short simulation times and small systems. Machine learning potentials have recently emerged as a third approach to model atomic interactions, and are purported to offer the accuracy of ab initio simulations with the speed of classical potentials. However, the performance of machine learning potentials depends crucially on the description of a local atomic environment. A set of invariant, orthogonal and differentiable descriptors for an atomic environment is proposed, implemented in a neural network potential for solid-state silicon, and tested in molecular dynamics simulations. Neural networks using the proposed descriptors are found to outperform ones using the Behler--Parinello and SOAP descriptors currently in the literature.
\end{abstract}

\maketitle

\section{Introduction}
\label{sec:intro}

Molecular dynamics (MD) simulations are frequently used in computational materials science to study the behaviour of both molecular and bulk systems. These simulations assume that the energy of an atom can be defined as a function of the local atomic environment, and the way this is done can dramatically affect the accuracy and performance of the simulation. The two main approaches in the literature are to calculate the atomic energies and forces using electronic structure calculations \cite{hautier2012computer}, resulting in \textit{ab initio} MD, or to pre-define functions describing the atomic interactions \cite{subramanian2009use}, resulting in \textit{classical} MD.  Perhaps the most popular electronic structure method in the literature is density functional theory (DFT) \cite{burke2012perspective} due to its relative accuracy for condensed matter states. However, the accuracy provided by DFT has a tremendous computational cost that strongly restricts the time and length scales of the simulation. In contrast to ab initio methods, the potentials used in classical MD are generally many orders of magnitude faster to execute. This makes them suitable for longer simulations containing many millions of atoms, allowing the study of more complex phenomena in larger domains. The drawback of such potentials is that they contain a limited number of fitting parameters that are generally calibrated to reproduce the properties of a particular bulk phase, resulting in inaccuracies when simulating complex phenomena such as phase transitions \cite{binks1993incorporation}, dislocations \cite{botu2016machine}, and interfacial dynamics \cite{sule2014classical}.

Recent interest in machine learning (ML) has encouraged the development of machine learning potentials (MLPs) with the goal of achieving quantum mechanical accuracy while approaching the speed of analytical potentials \cite{behler2007generalized,bartok2010gaussian,li2015molecular,deringer2017machine, ceriotti2018machine,bose2018machine}. These effectively apply non-parametric function regression to some reference data set to interpolate the potential energy surface (PES) of a local chemical environment. After a training process, the MLP is able to predict the energy of and force on a central atom from a description of the atomic neighborhood. Since ML algorithms can in principle reproduce even subtle many-body relationships, they provide higher flexibility than empirical potentials with a fixed functional form. Moreover, once they are trained on data collected by high-accuracy DFT simulations of a variety of configurations and phases, they can (given suitable coverage of the training data) maintain comparable accuracy during MD simulations with less computational expense than ab initio MD; they have been reported to be up to five orders of magnitude faster than quantum mechanical simulations with comparable accuracy \cite{behler2011neural,snyder2012finding,smith2017ani}.

Typical reference data to train a MLP consists of a central atom's local chemical information (relative positions and species of neighboring atoms) and potential energy. Similar to analytical potentials, the potential energy is assumed to depend only on the environment within some cutoff radius, and neighbors outside this volume are ignored. This atom-centered approach was initially proposed by Behler--Parinello \cite{behler2007generalized}, and enables one to calculate the total energy of a given system by summing over all the atoms. The preparation of the training data is crucial for an accurate representation of the PES, and DFT simulations are usually employed to calculate target energies and forces with high accuracy. Considering the cost of these DFT simulations, preparation of the training data is the most computationally demanding part of MLP development.

There are two key steps in the construction of a suitable reference set. First, since ML algorithms do not extrapolate as well as they interpolate, the space of local atomic environments should be widely sampled to increase the transferability of the potential. Several procedures have been proposed to construct the set of reference points used in training. 
Pukrittayakamee et al.\ \cite{pukrittayakamee2009simultaneous} used an importance sampling technique that selects training configurations based on atomic accelerations in MD simulations. This increases the sampling frequency where the potential energy gradient is large, i.e., where the PES is rapidly changing and could otherwise be sparsely sampled. Behler et al.\ \cite{behler2008pressure} attempted to equitably sample different regions of the configuration space by constructing a training set that included a mixture of crystal structures with different lattice parameters, amorphous structures, and some structures derived from metadynamics simulations. Another sampling technique that increases both the validity and accuracy of neural network potentials (NNPs) is extending the training set iteratively in a self-consistent way by detecting regions on the PES where the NNP performs poorly. Raff et al.\ \cite{raff2005ab} employed a primitive NNP in MD simulations to produce new trajectories. Energies associated with these new trajectories were then calculated with an ab initio method. Configurations from trajectories where contradictions occured were added to the reference set, and a new NNP was trained. The procedure was initialized with an empirical potential to obtain the first target energies, but the overall method was shown to be independent of the initial potential. They also claimed that most of the points in the configuration space are redundant and only a small subset of possible configurations needs to be sampled, and devised a novelty sampling algorithm to compute a set of possible trajectories in MD simulations. Behler suggested that multiple NNs could be used to find poorly represented regions on the PES by identifying regions where they conflicted, and appending these configurations to the training set \cite{behler2011neural}. Both iterative approaches were found to enhance the performance of a given NNP, and can be employed in conjunction.

The second major requirement for an MLP is some description of the local chemical environment as a set of real-valued numbers known as \textit{descriptors}. Machine learning algorithms are unaware of the physical properties of the data by design, and training can be dramatically simplified by appropriate pre-conditioning of the inputs. For an MLP, the description of the local chemical environment should be invariant with respect to fundamental physical symmetries including translations and rotations of the coordinate system and permutations of the atomic labels. If invariance to these symmetries is not enforced during the construction of the descriptors, the MLP could predict that physically identical configurations have different energies.
Let $\{ \vec{R}^{i1}, \vec{R}^{i2},...,\vec{R}^{iN} \}$ be the relative positions of the neighbors around the $i$th atom. These are usually converted into a vector of real-valued numbers $\{ G_1^i,G_2^i,...,G_{N_d}^i \}$ that are invariant to the physical symmetries. A recent review of MLPs and local structural descriptors by Behler \cite{behler2016perspective} included an overview of the Behler--Parinello (BP) symmetry functions \cite{behler2007generalized}, one of the first sets of descriptors proposed and widely used in the literature \cite{behler2008pressure, artrith2012high,morawietz2016van,natarajan2016neural}. Separately, Bart{\'o}k et al.\ \cite{bartok2013representing} reviewed several descriptors commonly used in the literature, proposed the SOAP descriptors to represent atomic environments, and quantitatively compared several variations using ad-hoc tests. The SOAP descriptors have increasingly been used in the literature, both within \cite{deringer2017machine,dragoni2018achieving} and without \cite{de2016comparing,rosenbrock2017discovering} the context of MLPs.

Since the literature on MLPs is relatively immature, there remains the possibility that local structural desciptors could be further improved. This study proposes a set of local structural descriptors that are found to contain considerably more information than the BP descriptors, and to be considerably more efficient to evaluate than the SOAP descriptors. The proposed descriptors were integrated into a NNP for solid-state silicon which was implemented as a new pair-style for LAMMPS \cite{plimpton1995fast} and validated in MD simulations. Since the main subject of this study is the descriptors rather than the potential, the energies of configurations in the reference set were calculated by means of an empirical potential \cite{stillinger1985computer} to reduce the computational cost. These would usually have been calculated with ab initio methods to achieve higher accuracy, but at the price of more uncertainty in the systematic error. Finally, the performance of the NNP using the proposed descriptors was compared with that of comparable NNPs using the BP and SOAP descriptors.

\section{Method}
\subsection{Descriptors}
\label{descriptors}

The faithfulness of any MLP depends strongly on how accurately the local structural descriptors describe the atomic neighborhood. A robust description would ideally provide a one-to-one mapping (bijection) between atomic positions and descriptors up to the symmetries of the physical system. This section introduces a new set of local structural descriptors that are continuous, twice-differentiable and invariant to the physical symmetries identified in Section \ref{sec:intro}.

Many steps in the construction resemble those for the SOAP descriptors \cite{bartok2013representing}. The first step is to map the list of relative atomic coordinates to a neighbor density function
\begin{align*}
 \rho^k(\vec{r}) = \sum_{j} w^k_{j} \delta(\vec{r}-\vec{r}^{ij}) 
\end{align*}
for a central atom $i$, thereby handling any permutation symmetries. The summation is performed over all neighbors $j$ within a spherical region defined by the cutoff radius $r_c$, which realizes the physical assumption that atomic energies should depend only on the local environment. Since all configurations are atom-centered, the neighbor position vectors $\vec{r}^{ij}$ are defined relative to the central atom. The weight factor $w^k_j$ could be used to distinguish the $k$th species in a multi-component system, but for simplicity is set to one and the superscript on $\rho^k(\vec{r})$ is dropped in the following.

The second step is to project $\rho(\vec{r})$ onto a set of orthonormal basis functions on the ball of radius $r_c$. Similar to Bart{\'o}k et al.\ \cite{bartok2013representing}, this projection is carried out by expanding $\rho(\vec{r})$ as
\begin{equation} \label{expansion}
\rho(\vec{r}) \approx \sum_{n = 0}^{n_\mathrm{max}} \sum_{l = 0}^{l_\mathrm{max}} \sum_{m = -l}^{l} c_{nlm} g_{n}(r) Y_{lm}(\theta,\phi)
\end{equation}
where $g_n(r)$ is a radial basis function, $Y_{lm}(\theta,\phi)$ is a spherical harmonic, and $n_\mathrm{max}$ and $l_\mathrm{max}$ are hyperparameters specifying the respective radial and angular resolutions. Although orthogonal radial basis functions should be preferred to minimize redundant information, Bart{\'o}k et al.\ \cite{bartok2013representing} neglected the appropriate weight factor for the spherical coordinate system and did not select orthogonal radial basis functions for their SO(3) and bispectrum descriptors, perhaps explaining the poor performance of these descriptors in their numerical experiments. Subsequent publications involving the SOAP descriptors \cite{szlachta2014accuracy} do use orthogonal radial basis functions, but this point is discussed further in Section \ref{sec:soap_comparison}. Apart from orthogonality, the radial basis functions should be defined to have vanishing values and first and second derivatives at the cutoff to ensure continuity of forces and elastic properties \cite{zhou2011effects}. Motivated by these requirements, we propose a set of radial basis functions constructed from linear combinations of the spherical Bessel functions.

Let $f_{nl}(r)$ be the linear combination of spherical Bessel functions
\begin{equation} \label{fnl}
f_{nl}(r)=a_{nl}j_l\left (r \frac{u_{nl}}{r_c}\right )+b_{nl}j_l\left (r \frac{u_{n+1,l}}{r_c}\right )
\end{equation}
where $a_{nl}$ and $b_{nl}$ are constants, $j_l(r)$ is the $l$th spherical Bessel function of the first kind, $u_{nl}$ is the $n$th root of $j_l(r)$, and $r_c$ is the cutoff. Since $f_{nl}(r_c)=0$ by definition, the objective is to find $a_{nl}$ and $b_{nl}$ such that $f'_{nl}(r_c)=0$ and $f''_{nl}(r_c)=0$. Combining the two differentiation rules for spherical Bessel functions in the Supplementary Material (SM) and solving for the roots of first and second derivatives indicates that both conditions can be satisfied if the coefficients in Eq.\ \ref{fnl} satisfy 
\begin{align*} 
a_{nl} &= \frac{u_{n+1,l}}{j_{l+1}(u_{nl})}c_{nl} \\
b_{nl} &= -\frac{u_{nl}}{j_{l+1}(u_{n+1,l})}c_{nl}
\end{align*}
for an arbitrary multiplicative constant $c_{nl}$. The value of $c_{nl}$ is fixed by requiring that the $f_{nl}(r)$ be normalized with respect to the inner product, leading to
\begin{equation*} 
f_{nl}(r)= \bigg( \frac{1}{r_c^3} \frac{2}{u_{nl}^2 + u_{n+1,l}^2} \bigg)^{1/2} \left [\frac{u_{n+1,l}}{j_{l+1}(u_{nl})}j_l\left (r \frac{u_{nl}}{r_c}\right ) - \frac{u_{nl}}{j_{l+1}(u_{n+1,l})}j_l\left (r \frac{u_{n+1,l}}{r_c} \right )\right ]
\end{equation*}
as an explicit equation for the $f_{nl}(r)$. A set of orthonormal radial basis functions $g_{nl}(r)$ can then be defined by applying the Gram-Schmidt process to the $f_{nl}(r)$ for $l \leq n \leq n_{max}$, with details provided in the SM.

\begin{figure}
\center
\subfigure[]{%
	\label{subfig:gn_values}{%
		\includegraphics[width=0.45\textwidth]{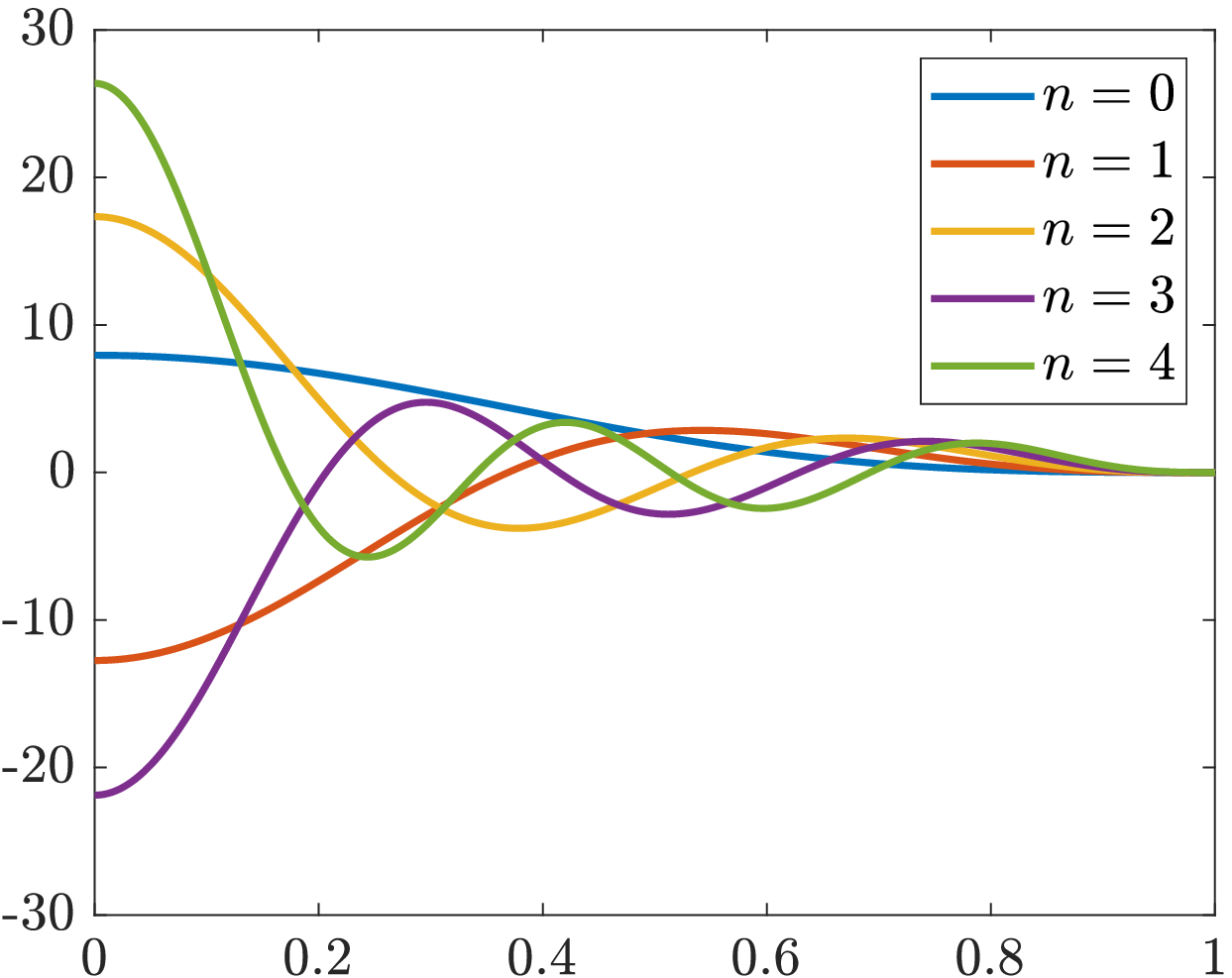}}} \hspace{0.5cm}
\subfigure[]{%
	\label{subfig:gn_fderiv}{%
		\includegraphics[width=0.46\textwidth]{%
			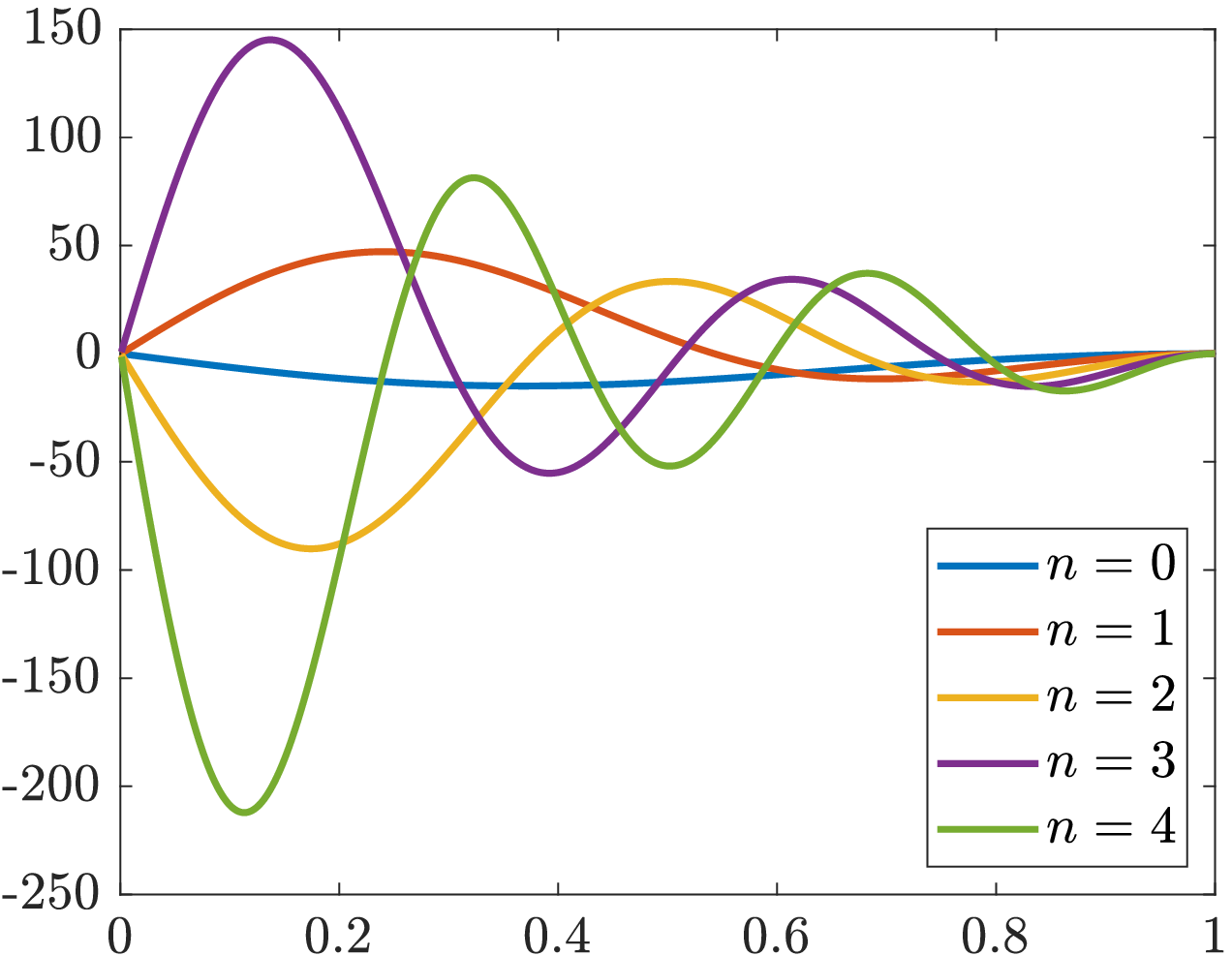}}}
\caption{The values (a) and the first derivatives (b) of the radial basis functions $g_{n}(r)$ for $0 \leq n \leq 4$ and $r_c = 1$. The behavior of the functions close to $r = r_c$ indicates that the second derivatives vanish there as well.}
\label{fig:radial_functions}
\end{figure}

Observing that $j_0(r) = \mathrm{sinc}(r)$ and $u_{n0} = (n + 1) \pi$, the evaluation of the radial basis functions simplifies considerably for the $l = 0$ case. The equation for $f_{n}(r) = f_{n0}(r)$ reduces to
\begin{equation*}
f_{n}(r)= (-1)^n \frac{\sqrt{2} \pi}{r_c^{3/2}} \frac{(n + 1) (n + 2)}{\sqrt{(n + 1)^2 + (n + 2)^2}} \left\{ \mathrm{sinc} \bigg[ r \frac{(n + 1) \pi}{r_c} \bigg] + \mathrm{sinc} \bigg[ r \frac{(n + 2) \pi}{r_c} \bigg] \right\}.
\end{equation*}
The radial basis functions $g_{n}(r) = g_{n0}(r)$ can then be defined by the recursion relations
\begin{align*}
e_{n} &= \frac{n^2 (n + 2)^2}{4 (n + 1)^4 + 1} \\
d_{n} &= 1 - \frac{e_{n}}{d_{n-1}} \\
g_{n}(r) &= \frac{1}{\sqrt{d_{n}}} \bigg[ f_{n}(r) + \sqrt{\frac{e_{n}}{d_{n-1}}} g_{n-1}(r) \bigg],
\end{align*}
initialized with $d_0 = 1$ and $g_0(r) = f_0(r)$. By construction, they satisfy the orthonormality condition
\begin{equation*}
\int_{0}^{r_c}g_{n'}(r)g_{n}(r)r^2\text{d}r = \delta_{n'n}
\end{equation*}
appropriate for functions on the ball of radius $r_c$. Several examples of the $g_{n}(r)$ and their first derivatives are shown in Fig.\ \ref{fig:radial_functions}.

With a suitable set of orthonormal basis functions defined on the ball of radius $r_c$, the expansion coefficients $c_{nlm}$ in Eq.\ \ref{expansion} can be written in terms of the relative spherical coordinates $(r^{ij},\theta^{ij},\phi^{ij})$ of the neighbors of the $i$th atom:
\begin{equation*} 
c_{nlm}=\sum_{j}^{}g_{n}(r^{ij})Y^*_{lm}(\theta^{ij},\phi^{ij}).
\end{equation*}
While the $c_{nlm}$ depend on the orientation of the coordinate system, the power spectrum $p_{nl}$ obtained from 
\begin{align*}
p_{nl}=\sum_{m=-l}^{l} c^{*}_{nlm} c_{nlm}
\end{align*}
is rotationally invariant \cite{bartok2013representing}. We therefore propose to use the real-valued $p_{nl}$ as local structural descriptors for neural networks. The number of descriptors is $(n_\mathrm{max} + 1) (l_\mathrm{max} + 1)$, and the accuracy of the expansion in Eq.\ \ref{expansion} increases with $n_\mathrm{max}$ and $l_\mathrm{max}$. That is, larger values of $n_\mathrm{max}$ and $l_\mathrm{max}$ include more terms in the approximation and more precisely specify the local environment. On the other hand, increasing the number of descriptors increases the cost of evaluating the NNP. While a local environment with $\nu > 1$ neighboring atoms requires precisely $3\nu - 3$ descriptors to describe the relative positions of all the atoms, we observe that more descriptors are often required in practice.

\subsection{Neural Network Potential}
\label{neuralnet}

Artificial Neural Networks (ANNs) have experienced a significant surge of interest in the last two decades after their success in various classification and regression problems \cite{hinton2012deep,yegnanarayana2009artificial,lecun2015deep}. In principle, NNs obey a universality theorem in that they are theoretically capable of reproducing any nonlinear functional relationship \cite{Csaji:Thesis:2001}. This encouraged their use for fitting PESs, where complex nonlinear relationships can exist between atomic configurations and atomic energies. While several different procedures have been proposed to develop NNPs                 \cite{behler2011neural,bholoa2007new,lorenz2004representing,blank1995neural}, the Behler--Parinello construction \cite{behler2007generalized} is followed here. The total energy $E$ of the system is decomposed into a sum of atomic contributions $E^i$:
\begin{equation*} 
E=\sum_{i} E^i.
\end{equation*}
Each atomic energy is calculated from the local chemical environment by an atomic NN. This atom-centered approach enables the modeling of systems with a variable number of atoms, overcoming a limitation of early NNs in the chemistry literature \cite{prudente1998fitting,bittencourt2004fitting,brown1996combining}.

\begin{figure}
\center
\includegraphics[width=7cm]{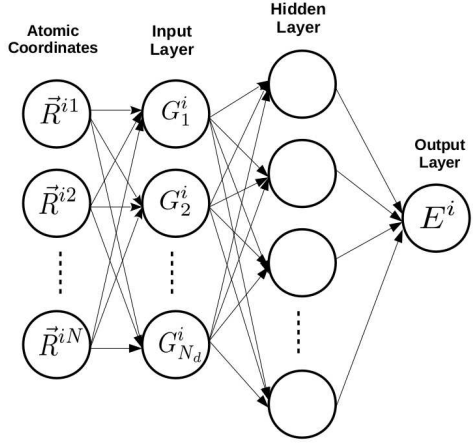}
\caption{Feed-forward neural network scheme used in this study. $E^i$ is the atomic potential energy of the $i$th atom, $G^i$ are the descriptors of the local environment, $\vec{R}^{ij}$ are the relative position vectors of the neighbors, $N$ is the number of neighbors and $N_d$ is the number of descriptors.}
\label{ffnn}
\end{figure}

The type of NNs that is generally used for fitting PESs is a \textit{feed-forward neural network} (FFNN) \cite{bebis1994feed} in which information only passes in a single direction towards the output layer. There is one input layer that feeds the relative atomic positions into the network and one output layer containing the atomic potential energy $E^i$. Some number of intervening hidden layers actually perform the regression, and the number of layers and the number of neurons in each layer are empirically optimized for the intended application. An example of a FFNN with one hidden layer is presented in Fig.\ \ref{ffnn}.

Let the hyperbolic tangent $h(x) = \tanh(x)$ be the transfer function for the hidden layers. The argument of the $j$th neuron in the first hidden layer is 
\begin{equation*}
a_j^1=b_j^1+\sum_{k=1}^{N_d} w_{kj}^1 G_{k},
\end{equation*}
the argument of the $j$th neuron in the $n$th hidden layer is 
\begin{equation*} 
a_j^n=b_j^n+\sum_{k} w_{kj}^n h(a_k^{n-1}),
\end{equation*}
and the value for the output neuron is
\begin{equation*} 
E^i=b_1^{N_L}+\sum_{k} w_{k1}^{N_L} h(a_k^{N_L-1}),
\end{equation*}
where $N_L$ is the number of layers, $w_{kj}^n$ is the weight that binds the $j$th neuron in the $n$th layer to the $k$th neuron in the $(n-1)$th layer, and $b_j^n$ is the bias for the $j$th neuron of the $n$th layer. The weights and biases constitute the parameter space to be fitted during the training of the NN.

The number of hidden layers is of great importance and can dramatically affect both the accuracy and performance of MD simulations. Additional layers enhance the ability of the NN to fit complex functions, but have the drawback of increasing the number of weights and biases to optimize, possibly slowing down or even hindering the training process \cite{hochreiter1998vanishing}. Redundant layers and neurons can also cause over-fitting, meaning that the NN becomes less capable of extrapolating to configurations outside of the training set. In regression problems such as PES fitting, this can be a severe problem and significantly reduce the reliability of the NNP in energy and force predictions. Therefore, it is preferable to use the smallest possible number of layers and neurons that achieve the desired error when building the NN. We decided to use a single hidden layer after observing that additional layers did not substantially improve the fitting accuracy.

The NNs were trained using the standard back-propogation \cite{rumelhart1986learning} and stochastic gradient descent \cite{robbins1985stochastic} algorithms. The root mean square error (RMSE) 
\begin{equation*} 
\Gamma = \bigg [ \frac{1}{N_T}\sum_{i=1}^{N_T}(E_\mathrm{pre}^i-E_\mathrm{act}^i)^2 \bigg ]^{1/2}
\end{equation*}
was used to quantify the error after each epoch, where $N_T$ is the total number of training points and $E_{pre}^i$ and $E_{act}^i$ are the predicted and actual potential energies, respectively. The mini-batch size was 100 for all simulations. All of the training processes were performed in Python using Keras with the TensorFlow backend \cite{chollet2015keras,abadi2016tensorflow}.

Given the atomic energies, the forces acting on each atom can be computed from the gradient of $E$. This requires repeated application of the chain rule due to the dependence of the descriptors on the atomic positions. Let the $j$th component of the force on the $i$th atom be $F_{j}^i$. This is obtained by summing the contributions from all $N$ atoms in the system by
\begin{equation*}
F_{j}^i = -\sum_{k=1}^{N} \frac{\partial E^k}{\partial r_{j}^i}= -\sum_{k=1}^{N}\sum_{p=1}^{N_d}\frac{\partial E^k}{\partial G_{p}^k}\frac{\partial G_{p}^k}{\partial r_{j}^i},
\end{equation*}
where $r^{i}_{j}$ is the $j$th Cartesian coordinate of the $i$th atom, $G^k_p$ is the $p$th descriptor for the $k$th atom, and $N_d$ is the number of descriptors. The derivatives $\partial E^k/\partial G_{p}^k$ depend only on the NN architecture and can be calculated by back-propogation. The derivatives $\partial G_{p}^k/ \partial r_{j}^i$ of the proposed descriptors with respect to the Cartesian coordinates and other details of the force calculation are provided in the SM. Note that the force includes contributions from the dependence of the neighboring atoms' energies on the position of the $i$th atom, and from the dependence of the energy of the $i$th atom on its own position---displacing the $i$th atom by $\Delta \vec{r}$ effectively displaces the surrounding atoms by -$\Delta \vec{r}$, contributing to the total force.

\subsection{Training Data}
\label{refset}

The training data set should generally be prepared carefully, as the selection of configurations to include can significantly affect the performance and accuracy of the NN. If an atomic configuration that is not adequately represented in the reference set occurs during simulation, the error in the predicted potential energy could increase dramatically. This can be addressed by directly sampling points in diverse regions of the configuration space, e.g., by considering all possible structures represented on the phase diagram \cite{behler2008pressure}, but there is no guarantee that other configurations would not occur in simulation. A second option would be to employ an importance sampling method to enhance the flexibility and extrapolation capability of NNs. Since our main intention is to investigate the properties of the proposed descriptors rather than develop a general-purpose NNP, training configurations were only sampled from MD simulations of silicon within a limited temperature range. Sampling was performed using the algorithm proposed by Pukrittayakamee et al.\ \cite{pukrittayakamee2009simultaneous} and modified by Stende \cite{Stende:Thesis:2017}, which was observed to reduce the fitting error. The algorithm consists of sampling the local environment around a given atom at a variable interval
\begin{equation*} 
\tau = \begin{cases}
 1 & |\bar{F}^i|>\alpha \\
 \left \lfloor \alpha/|\bar{F}^i| \right \rfloor & |\bar{F}^i| \leq\alpha \\
 \tau_\mathrm{max} & \left \lfloor \alpha/|\bar{F}^i| \right \rfloor >\tau_\mathrm{max}
\end{cases}
\end{equation*}
where $\bar{F}^i$ is the total force acting on the $i$th atom and $\tau$ is measured in units of the MD timestep. We tracked {\small $\sim$}10 atoms throughout an MD simulation using the Stillinger--Weber potential \cite{stillinger1985computer}, calcualted the corresponding forces, and sampled training set configurations at intervals specified by $\tau$. The inverse relationship between $|\bar{F}^i|$ and $\tau$ ensures that high-gradient regions on the PES are more equitably represented in the training data, and is observed to reduce the fitting error. Note that $\tau_\mathrm{max}$ and $\alpha$ are system-dependent parameters.

\section{Results and Discussion}

The performance of the descriptors proposed in Section \ref{descriptors} in an NNP for solid-state silicon is compared with that of the BP descriptors and the SOAP descriptors. The NNP is further validated by measuring the elastic constants of solid-state silicon. The Stillinger--Weber potential \cite{stillinger1985computer} is selected as the ground truth, and was used to calculate the energies of all configurations in the training set.

\subsection{Behler--Parinello Descriptors}

\begin{figure}
\center
\includegraphics[width=8cm]{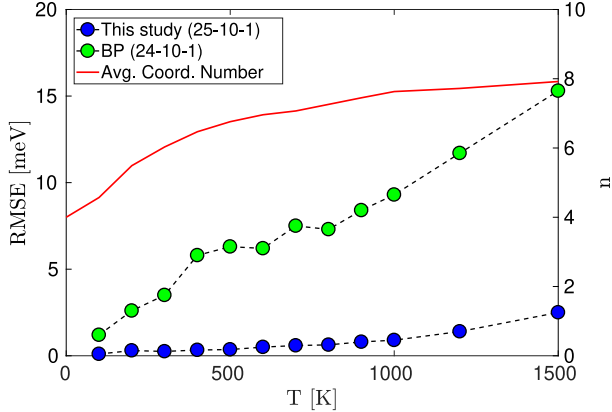}
\caption{Performance of the BP descriptors and the proposed descriptors with increasing temperature. The left and right y-axis show the RMSE in meV and the average number of neighbors $n$, respectively.}
\label{compare}
\end{figure}

The BP descriptors were one of the earliest sets of descriptors used for MLPs, and are still used for this purpose. Following Behler \cite{behler2007generalized}, the radial symmetry functions $G_i^r$ and angular symmetry functions $G_i^a$ are defined as 
\begin{align*} 
G_i^r &= \sum_{j=1}^{N}e^{-\eta(r_{ij}-r_s)^2}f_c(r_{ij}) \\
G_i^a &= 2^{1-\zeta}\sum_{j\neq i}\sum_{k>j}[(1+\lambda \text{cos}\theta_{jik})^{\zeta}e^{-\eta(r_{ij}^2 + r_{ik}^2)}f_c(r_{ij})f_c(r_{ik})]
\end{align*}
where $f_c(r_{ij})$ is a cutoff function and $\eta$, $\lambda$, $\zeta$ and $r_s$ are adjustable parameters. These functions are designed to create a set of real-valued numbers from the atomic distances $r_{ij}$ and bond angles $\theta_{ijk}$. A recent study \cite{Stende:Thesis:2017} developed a single hidden layer NNP for silicon using the BP descriptors and a 24-10-1 architecture. The performance of this NNP is compared to one using our descriptors with a 25-10-1 architecture (the number of descriptors is not exactly the same because of indexing). The RMSE for the validation set was evaluated as a function of sampling temperature while keeping the training set and all other hyperparameters fixed. The results in Fig.\ \ref{compare} suggest that our descriptors provide considerably more information about the local environment and result in a more accurate NNP than the BP descriptors. Alternatively, considerably fewer of our descriptors would be required to construct an NNP of a given accuracy, reducing the expense of force calculations in MD simulations. Moreover, the proposed descriptors contain few adjustable parameters ($n_\mathrm{max}$, $l_\mathrm{max}$ and $r_c$) and should therefore be widely applicable with minimal calibration, whereas the parameters $\eta$, $\lambda$, $\zeta$ and $r_s$ need to be adjusted for the BP descriptors. 

\begin{figure}
\center
\includegraphics[width=8cm]{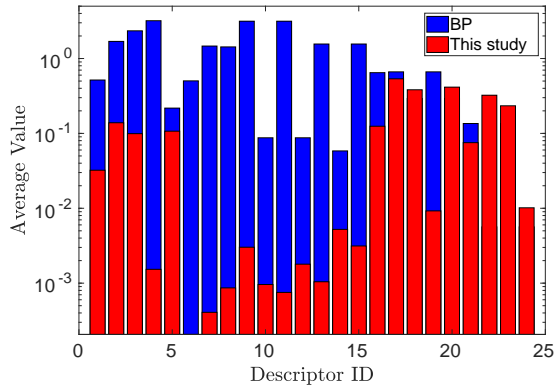}
\caption{Average values of the proposed descriptors and the BP descriptors for a single training data set consisting of $10^4$ silicon configurations at 300 K.}
\label{compare2}
\end{figure}

We also observed that NNs using the BP descriptors were more difficult to train than ones using our descriptors. Similar to other ML algorithms, NNs often require detailed pre-processing of the input data to obtain reasonable results. One frequent problem is saturation of some hidden neurons during training, resulting in trapping around a local minimum that prevents further learning. This is related to the vanishing gradient problem, which is one of the most common issues with artificial neural networks and happens more frequently when some input values are much larger than others. The wide variation in the magnitudes of the BP descriptors, as indicated by Fig.\ \ref{compare2}, likely caused the observed difficulties with training. While Behler suggested several pre-processing techniques to overcome this issue \cite{behler2011neural}, we found that the problem could be solved by initializing the weight matrices with values from the Xavier normal distribution \cite{glorot2010understanding} with a variance of $\sqrt{6 /(n_{l-1} + n_l)}$ where $n_l$ is the number of neurons in the $l$th layer. By contrast, training with our descriptors progressed the same regardless of the weight initialization and without any additional pre-processing. The only advantage of the BP descriptors we observed was that they required roughly half the time to evaluate (with our naive implementations), but this seems to be strongly outweighted by the advantage in accuracy.

\subsection{SOAP Descriptors}
\label{sec:soap_comparison}

Bart{\'o}k, Kondor, and Cs{\'a}nyi initially introduced the Smooth Overlap of Atomic Positions (SOAP) descriptors \cite{bartok2013representing} to support modeling the PES as a Gaussian process \cite{bartok2010gaussian}. Rather than directly using the SOAP descriptors as inputs into an MLP though, an inner product of normalized descriptor vectors (the SOAP kernel) is generally employed to measure the similarity of a pair of atomic environments. If $\vec{G}_i$ is the vector of SOAP descriptors for the $i$th atom, the SOAP kernel $K_{ij}$ comparing the environments around the $i$th and $j$th atoms is defined by means of\cite{szlachta2014accuracy} 
\begin{align*}
\nvec{G}_i &= \vec{G}_i / | \vec{G}_i | \\
K_{ij} &= \sigma_w^2 | \nvec{G}_i \cdot \nvec{G}_j |^\xi
\end{align*}
where $\sigma_w$ and $\xi$ are adjustable parameters. The quantity $d_{ij} = \sqrt{1 - K_{ij}}$ has been said to be a metric \cite{de2016comparing}, though the identity property of a metric requires that the distance vanish if and only if the configurations around the $i$th and $j$th atoms are identical. Consider the case where, for every atom in the neighborhood of the $i$th atom, there is a corresponding pair of atoms separated by an arbitrarily small distance $\delta$ in the same position relative to the $j$th atom. The expansion coefficients $c_{nlm}$ for the two configurations would then differ by roughly a factor of two, the SOAP descriptors $\vec{G}_i$ by roughly a factor of four, and the distance $d_{ij}$ could be made arbitrarily close to zero by adjusting the value of $\delta$. That is, the quantity $d_{ij}$ does not satisfy the identity property and is not a metric. The difficulty seems to be essential in that, if the vectors of SOAP descriptors were not normalized, the magnitude of $K_{ij}$ would not be bounded above, and the value for which environments are considered similar would no longer be unique. The existence of this counterexample does little to inspire confidence that there are not others, particularly since this is a function in a high-dimensional space where intuition is difficult to develop.

Instead of the SOAP kernel, this study uses the SOAP descriptors as inputs for an NNP. The derivation of the proposed descriptors is closely related to that of the SOAP descriptors in several respects; a neighbor density function is projected onto a set of orthogonal basis functions, and the descriptors are given by inner products of vectors of the expansion coefficients. That said, there are several significant differences. First, the neighbor density function for the proposed descriptors is a sum of Dirac delta functions, whereas that for the SOAP descriptors is a sum of Gaussians. This has the consequence that evaluating the SOAP descriptors involves a relatively expensive numerical integration, whereas the proposed descriptors can be found merely by evaluating the relevant basis functions at the neighboring atoms' positions. Using a sum of Gaussians is said to improve the stability of the SOAP descriptors with respect to perturbations of the atoms' positions \cite{bartok2013representing}, but the differentiability of the basis functions in Section \ref{descriptors} is sufficient to give the proposed descriptors the same property. Second, the SOAP descriptors are given by the inner products of vectors of expansion coefficient with different values of $n$:
\begin{equation*}
p_{n'nl} = \sum_m c_{n'lm}^* c_{nlm}.
\end{equation*}
Depending on the radial basis functions this could help to couple information from different spherical shells within the domain, but increases the number of descriptors and the computational expense of evaluating an NNP for fixed $n_\mathrm{max}$ and $l_\mathrm{max}$. The proposed descriptors instead depend on a set of orthonormal basis functions with strong radial overlap (visible in Fig.\ \ref{subfig:gn_values}), obviating the need for such explicit coupling.

\begin{figure}
\center
\includegraphics[width=8cm]{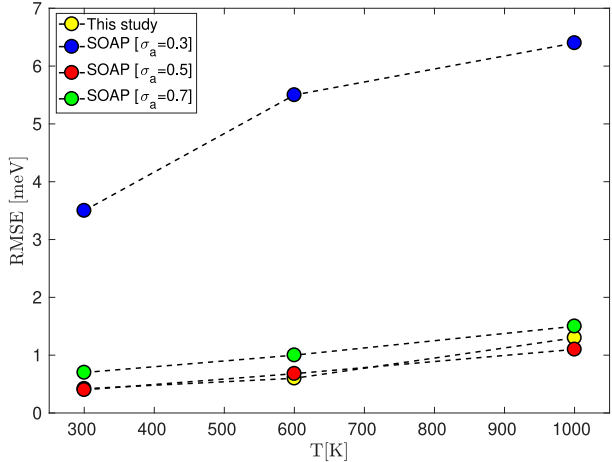}
\caption{Performance of the SOAP descriptors and the proposed descriptors with increasing temperature, with NN architectures of (18-10-1) and (16-10-1), respectively. $\sigma_a$ is the standard deviation of the Gaussians used to generate the neighbor density function in angstroms. All other fitting parameters for the SOAP descriptors were taken from the literature \cite{szlachta2014accuracy}.}
\label{fig:soap_compare}
\end{figure}

Evidence that these differences do not degrade the performance of the proposed descriptors relative to the SOAP descriptors is given in Fig.\ \ref{fig:soap_compare}. The performance of an NNP trained with $16$ of the proposed descriptors is nearly identical to that of an NNP trained with $18$ of the SOAP descriptors for the optimal value of the Guassian width in the neighbor density function. Additionally, the proposed descriptors have several advantages that are not visible from this figure. First, computing $16$ of the proposed descriptors for $100$ training points requires {\small $\sim$}$0.2$ seconds whereas computing $18$ of the SOAP descriptors requires {\small $\sim$}$9.5$ seconds (with our naive implementations). The special function evaluations and numerical integrations required for the SOAP descriptors would likely be expensive even in optimized code. Second, the second derviatives of the proposed descriptors are continuous to atoms passing through the domain boundary, whereas only the first derivatives are continuous for the SOAP descriptors \cite{szlachta2014accuracy}. This is significant because discontinuous second derivatives of the potential energy have been observed to lead to discontinuous elastic constants and anomalous thermal transport in MD simulations \cite{zhou2011effects}. Third, the proposed descriptors do not require an arbitrary choice of cutoff function, and the number of adjustable parameters is smaller than for the SOAP descriptors. Specifically, the proposed descriptors only require that $r_c$, $n_\mathrm{max}$, and $l_\mathrm{max}$ be specified, whereas the SOAP descriptors have up to six adjustable parameters \cite{szlachta2014accuracy} if the Gaussian widths in the neighbor density function and in the raw radial basis functions are allowed to be independent. Setting these adjustable parameters introduces additional complexity, with Fig.\ \ref{fig:soap_compare} showing the sensitivity of NNP performance to the value of one of them.

\subsection{NNP Validations}

\begin{table}
\centering
\small
\caption{The minimum RMSE per atom for different temperatures $T$ and NN architectures, where $n$ is the average number of neighbors and $N_d$ is the number of descriptors. All of the neural networks were trained on 8500 training points for 20000 epochs, and RMSE values were obtained on 1500 test configurations that are not included in the training set.\\} 
\begin{tabular}{c@{\hskip 0.33in}c@{\hskip 0.33in}c@{\hskip 0.33in}c@{\hskip 0.33in}c@{\hskip 0.33in}c@{\hskip 0.33in}c}
\hline
T [K] & $n$ & NN & $N_d$  & RMSE [meV] \\
 \hline   
300 & 6.03 & 16-8-1 & 16  & 0.22\\  
300 & 6.03 & 16-16-1 & 16  & 0.23 \\
300 & 6.03 & 25-8-1 & 25   & 0.35 \\
600 & 6.96 & 16-8-1 & 16  & 0.56 \\  
600 & 6.96 & 16-16-1 & 16 & 0.64 \\ 
600 & 6.96 & 25-8-1 & 25 & 0.51 \\ 
1000 & 7.63 & 16-8-1 & 16 & 1.24 \\ 
1000 & 7.63 & 16-16-1 & 16 & 1.98 \\  
1000 & 7.63 & 25-8-1 & 25 & 0.88 \\     
1500 & 7.92 & 16-8-1 & 16 & 2.62\\
1500 & 7.92 & 16-16-1 & 16 & 2.75\\  
1500 & 7.92 & 25-8-1 & 25 & 2.3 \\   \hline
\end{tabular} \label{table1}
\end{table}

We further investigated the performance of NNPs using our descriptors at a variety of sampling temperatures and NN architectures, with the results reported in Table \ref{table1}. The number of accessible configurations in an MD simulation usually increases rapidly with temperature, meaning that any given accuracy would require more descriptors to encode the neighborhoods and training points to cover the configuration space. The parameters $n_\mathrm{max}$ and $l_\mathrm{max}$ in Eq.\ \ref{expansion} set the number of descriptors, with higher values resulting in more terms in the approximation of the neighbor density function and generally lower fitting errors. The average number of neighbors $n$ varied from $6$ to $8$ within the selected temperature range, implying that a minimum of $15$ to $21$ descriptors were required. This is consistent with our observations that using more than $25$ descriptors ($n_\mathrm{max} = l_\mathrm{max} = 4$) did not substantially decrease the RMSE, and is consistent with the number of descriptors used in other NNP studies \cite{raff2005ab,behler2007generalized, behler2008pressure,artrith2012high}. Moreover, when the temperature was elevated to 1500 K (the melting point of silicon is 1687 K), NNs using $25$ descriptors consistently outperformed those using $16$ descriptors; the higher average number of neighbors at these temperatures allowed more complex configurations that required more descriptors.

Using more than one hidden layer or more than ten hidden neurons did not substantially improve the accuracy of the NNP. Table \ref{table1} indicates that using more hidden neurons actually \emph{decreased} the accuracy, perhaps as a consequence of the increased complexity of the training process. This differs from previous studies that used the BP descriptors in two-layer NNPs for single-species systems \cite{behler2007generalized,behler2008pressure,artrith2012high}. Artrith and Behler \cite{artrith2012high} further mentioned that monocomponent systems typically require $40$ to $60$ symmetry functions to achieve a complete description, and used $51$ in their study. The reason for the difference in behavior with that observed here is not known, but is conjectured to be related to our descriptors deriving from an efficient expansion of the neighbor density function using orthogonal basis functions, and to our radial basis functions effectively coupling information in multiple spherical shells.

\begin{table}
\centering
\small
\caption{Bulk modulus ($K$), shear modulus ($G$) and Poisson's ratio ($\nu$) of solid-state silicon at 300 K as measured in MD simulations using the analytic SW potential and our NNP.\\} 
\begin{tabular}{c@{\hskip 0.25in}c@{\hskip 0.25in}c@{\hskip 0.25in}c@{\hskip 0.25in}c@{\hskip 0.25in}c@{\hskip 0.25in}c}
\hline
 & K [GPa] & G [GPa] & $\nu$ \\
 \hline   
SW \cite{stillinger1985computer} & 101.4 & 56.4 & 0.335 \\
NNP & 101.7 $\pm 0.3$ & 56.3 $\pm 0.1$  & 0.337 $\pm 0.2$ \\ \hline
\end{tabular} \label{table2}
\end{table}

Finally, the NNP developed here was added as a new pair-style to LAMMPS. The bulk modulus, shear modulus and Poisson's ratio of solid-state silicon were calculated from an MD simulation using an NNP with our descriptors and a 25-10-1 architecture at 300 K, and compared with those reported for the SW potential. The results in Table \ref{table2} offer additional evidence that our NNP is able to reproduce features of the potential energy surface with excellent accuracy.

\section{Conclusions}

Belonging to the family of machine learning force-fields, high-dimensional neural network potentials have been found to be viable alternatives to electronic structure calculations by providing similar levels of accuracy at a lower computational cost. One crucial requirement for developing a robust neural network potential is a description of the local atomic neighborhood as a set of symmetrically-invariant real-valued numbers. Referred to as descriptors, different constructions have been proposed in the literature, but there is as yet no established canonical choice due to the recent emergence of the field. This paper introduces a new set of orthogonal descriptors that are invariant to the physical symmetries and can more efficiently represent structural environments than two of the frequent alternatives \cite{behler2007generalized,bartok2013representing}.

The performance of the proposed descriptors in a neural network potential was compared to that of the Behler--Parinello descriptors and the SOAP descriptors, both commonly employed in machine learning potentials. For a given training set and comparable hyperparameters, our descriptors were found to give substantially smaller fitting errors than the Behler--Parinello descriptors, and similar fitting errors to the SOAP descriptors but at an order of magnitude lower computational cost. The superior performance of the proposed descriptors as compared to the Behler--Parinello descriptors is conjectured to be a consequence of the proposed descriptors deriving from a function expansion over orthogonal basis functions that efficiently encodes configurational information. As for the SOAP descriptors, the improved computational efficiency is a consequence of avoiding special function evaluations and numerical integration. Finally, the suitability of the proposed descriptors for machine learning potentials was verified by preliminary molecular dynamics simulations of solid-state silicon.

\section{Supplementary Material}

The detailed derivation of the proposed radial basis functions and the force calculation procedure can be found in the Supplementary Material.


\bibliographystyle{unsrtnat}
\bibliography{refs.bib}

\begin{thebibliography}{48}
\providecommand{\natexlab}[1]{#1}
\providecommand{\url}[1]{\texttt{#1}}
\expandafter\ifx\csname urlstyle\endcsname\relax
  \providecommand{\doi}[1]{doi: #1}\else
  \providecommand{\doi}{doi: \begingroup \urlstyle{rm}\Url}\fi

\bibitem[Hautier et~al.(2012)Hautier, Jain, and Ong]{hautier2012computer}
Geoffroy Hautier, Anubhav Jain, and Shyue~Ping Ong.
\newblock {From the computer to the laboratory: materials discovery and design
  using first-principles calculations}.
\newblock \emph{Journal of Materials Science}, 47\penalty0 (21):\penalty0
  7317--7340, 2012.

\bibitem[Subramanian(2009)]{subramanian2009use}
Lalitha Subramanian.
\newblock {Use of Force Fields in Materials Modeling}.
\newblock \emph{Reviews in Computational Chemistry}, 16:\penalty0 141, 2009.

\bibitem[Burke(2012)]{burke2012perspective}
Kieron Burke.
\newblock {Perspective on density functional theory}.
\newblock \emph{The Journal of chemical physics}, 136\penalty0 (15):\penalty0
  150901, 2012.

\bibitem[Binks and Grimes(1993)]{binks1993incorporation}
D~Jason Binks and Robin~W Grimes.
\newblock {Incorporation of monovalent ions in ZnO and their influence on
  varistor degradation}.
\newblock \emph{Journal of the American Ceramic Society}, 76\penalty0
  (9):\penalty0 2370--2372, 1993.

\bibitem[Botu et~al.(2016)Botu, Batra, Chapman, and Ramprasad]{botu2016machine}
Venkatesh Botu, Rohit Batra, James Chapman, and Rampi Ramprasad.
\newblock {Machine learning force fields: construction, validation, and
  outlook}.
\newblock \emph{The Journal of Physical Chemistry C}, 121\penalty0
  (1):\penalty0 511--522, 2016.

\bibitem[S{\"u}le and Szendr\H{o}(2014)]{sule2014classical}
P{\'e}ter S{\"u}le and M~Szendr\H{o}.
\newblock {The classical molecular dynamics simulation of graphene on Ru (0001)
  using a fitted Tersoff interface potential}.
\newblock \emph{Surface and Interface Analysis}, 46\penalty0 (1):\penalty0
  42--47, 2014.

\bibitem[Behler and Parrinello(2007)]{behler2007generalized}
J{\"o}rg Behler and Michele Parrinello.
\newblock {Generalized neural-network representation of high-dimensional
  potential-energy surfaces}.
\newblock \emph{Physical review letters}, 98\penalty0 (14):\penalty0 146401,
  2007.

\bibitem[Bart{\'o}k et~al.(2010)Bart{\'o}k, Payne, Kondor, and
  Cs{\'a}nyi]{bartok2010gaussian}
Albert~P Bart{\'o}k, Mike~C Payne, Risi Kondor, and G{\'a}bor Cs{\'a}nyi.
\newblock {Gaussian approximation potentials: The accuracy of quantum
  mechanics, without the electrons}.
\newblock \emph{Physical review letters}, 104\penalty0 (13):\penalty0 136403,
  2010.

\bibitem[Li et~al.(2015)Li, Kermode, and {De Vita}]{li2015molecular}
Zhenwei Li, James~R Kermode, and Alessandro {De Vita}.
\newblock {Molecular dynamics with on-the-fly machine learning of
  quantum-mechanical forces}.
\newblock \emph{Physical review letters}, 114\penalty0 (9):\penalty0 096405,
  2015.

\bibitem[Deringer and Cs{\'a}nyi(2017)]{deringer2017machine}
Volker~L Deringer and G{\'a}bor Cs{\'a}nyi.
\newblock {Machine learning based interatomic potential for amorphous carbon}.
\newblock \emph{Physical Review B}, 95\penalty0 (9):\penalty0 094203, 2017.

\bibitem[Ceriotti et~al.(2018)Ceriotti, Willatt, and
  Cs{\'a}nyi]{ceriotti2018machine}
Michele Ceriotti, Michael~J Willatt, and G{\'a}bor Cs{\'a}nyi.
\newblock {Machine Learning of Atomic-Scale Properties Based on Physical
  Principles}.
\newblock \emph{Handbook of Materials Modeling: Methods: Theory and Modeling},
  pages 1--27, 2018.

\bibitem[Bose et~al.(2018)Bose, Dhawan, Nandi, Sarkar, and
  Ghosh]{bose2018machine}
Samik Bose, Diksha Dhawan, Sutanu Nandi, Ram~Rup Sarkar, and Debashree Ghosh.
\newblock {Machine learning prediction of interaction energies in rigid water
  clusters}.
\newblock \emph{Physical Chemistry Chemical Physics}, 20\penalty0
  (35):\penalty0 22987--22996, 2018.

\bibitem[Behler(2011)]{behler2011neural}
J{\"o}rg Behler.
\newblock {Neural network potential-energy surfaces in chemistry: a tool for
  large-scale simulations}.
\newblock \emph{Physical Chemistry Chemical Physics}, 13\penalty0
  (40):\penalty0 17930--17955, 2011.

\bibitem[Snyder et~al.(2012)Snyder, Rupp, Hansen, M{\"u}ller, and
  Burke]{snyder2012finding}
John~C Snyder, Matthias Rupp, Katja Hansen, Klaus-Robert M{\"u}ller, and Kieron
  Burke.
\newblock {Finding density functionals with machine learning}.
\newblock \emph{Physical review letters}, 108\penalty0 (25):\penalty0 253002,
  2012.

\bibitem[Smith et~al.(2017)Smith, Isayev, and Roitberg]{smith2017ani}
Justin~S Smith, Olexandr Isayev, and Adrian~E Roitberg.
\newblock {ANI-1: an extensible neural network potential with DFT accuracy at
  force field computational cost}.
\newblock \emph{Chemical science}, 8\penalty0 (4):\penalty0 3192--3203, 2017.

\bibitem[Pukrittayakamee et~al.(2009)Pukrittayakamee, Malshe, Hagan, Raff,
  Narulkar, Bukkapatnum, and Komanduri]{pukrittayakamee2009simultaneous}
A~Pukrittayakamee, M~Malshe, M~Hagan, LM~Raff, R~Narulkar, S~Bukkapatnum, and
  R~Komanduri.
\newblock {Simultaneous fitting of a potential-energy surface and its
  corresponding force fields using feedforward neural networks}.
\newblock \emph{The Journal of chemical physics}, 130\penalty0 (13):\penalty0
  134101, 2009.

\bibitem[Behler et~al.(2008)Behler, Marto\v{n}{\'a}k, Donadio, and
  Parrinello]{behler2008pressure}
J{\"o}rg Behler, Roman Marto\v{n}{\'a}k, Davide Donadio, and Michele
  Parrinello.
\newblock {Pressure-induced phase transitions in silicon studied by neural
  network-based metadynamics simulations}.
\newblock \emph{Physica status solidi (b)}, 245\penalty0 (12):\penalty0
  2618--2629, 2008.

\bibitem[Raff et~al.(2005)Raff, Malshe, Hagan, Doughan, Rockley, and
  Komanduri]{raff2005ab}
LM~Raff, M~Malshe, M~Hagan, DI~Doughan, MG~Rockley, and R~Komanduri.
\newblock {Ab initio potential-energy surfaces for complex, multichannel
  systems using modified novelty sampling and feedforward neural networks}.
\newblock \emph{The Journal of chemical physics}, 122\penalty0 (8):\penalty0
  084104, 2005.

\bibitem[Behler(2016)]{behler2016perspective}
J{\"o}rg Behler.
\newblock {Perspective: Machine learning potentials for atomistic simulations}.
\newblock \emph{The Journal of chemical physics}, 145\penalty0 (17):\penalty0
  170901, 2016.

\bibitem[Artrith and Behler(2012)]{artrith2012high}
Nongnuch Artrith and J{\"o}rg Behler.
\newblock {High-dimensional neural network potentials for metal surfaces: A
  prototype study for copper}.
\newblock \emph{Physical Review B}, 85\penalty0 (4):\penalty0 045439, 2012.

\bibitem[Morawietz et~al.(2016)Morawietz, Singraber, Dellago, and
  Behler]{morawietz2016van}
Tobias Morawietz, Andreas Singraber, Christoph Dellago, and J{\"o}rg Behler.
\newblock {How van der Waals interactions determine the unique properties of
  water}.
\newblock \emph{Proceedings of the National Academy of Sciences}, 113\penalty0
  (30):\penalty0 8368--8373, 2016.

\bibitem[Natarajan and Behler(2016)]{natarajan2016neural}
Suresh~Kondati Natarajan and J{\"o}rg Behler.
\newblock {Neural network molecular dynamics simulations of solid--liquid
  interfaces: Water at low-index copper surfaces}.
\newblock \emph{Physical Chemistry Chemical Physics}, 18\penalty0
  (41):\penalty0 28704--28725, 2016.

\bibitem[Bart{\'o}k et~al.(2013)Bart{\'o}k, Kondor, and
  Cs{\'a}nyi]{bartok2013representing}
Albert~P Bart{\'o}k, Risi Kondor, and G{\'a}bor Cs{\'a}nyi.
\newblock {On representing chemical environments}.
\newblock \emph{Physical Review B}, 87\penalty0 (18):\penalty0 184115, 2013.

\bibitem[Dragoni et~al.(2018)Dragoni, Daff, Cs{\'a}nyi, and
  Marzari]{dragoni2018achieving}
Daniele Dragoni, Thomas~D Daff, G{\'a}bor Cs{\'a}nyi, and Nicola Marzari.
\newblock {Achieving DFT accuracy with a machine-learning interatomic
  potential: Thermomechanics and defects in bcc ferromagnetic iron}.
\newblock \emph{Physical Review Materials}, 2\penalty0 (1):\penalty0 013808,
  2018.

\bibitem[De et~al.(2016)De, Bart{\'o}k, Cs{\'a}nyi, and
  Ceriotti]{de2016comparing}
Sandip De, Albert~P Bart{\'o}k, G{\'a}bor Cs{\'a}nyi, and Michele Ceriotti.
\newblock {Comparing molecules and solids across structural and alchemical
  space}.
\newblock \emph{Physical Chemistry Chemical Physics}, 18\penalty0
  (20):\penalty0 13754--13769, 2016.

\bibitem[Rosenbrock et~al.(2017)Rosenbrock, Homer, Cs{\'a}nyi, and
  Hart]{rosenbrock2017discovering}
Conrad~W Rosenbrock, Eric~R Homer, G{\'a}bor Cs{\'a}nyi, and Gus~LW Hart.
\newblock {Discovering the building blocks of atomic systems using machine
  learning: application to grain boundaries}.
\newblock \emph{npj Computational Materials}, 3\penalty0 (1):\penalty0 29,
  2017.

\bibitem[Plimpton(1995)]{plimpton1995fast}
Steve Plimpton.
\newblock {Fast parallel algorithms for short-range molecular dynamics}.
\newblock \emph{Journal of computational physics}, 117\penalty0 (1):\penalty0
  1--19, 1995.

\bibitem[Stillinger and Weber(1985)]{stillinger1985computer}
Frank~H Stillinger and Thomas~A Weber.
\newblock {Computer simulation of local order in condensed phases of silicon}.
\newblock \emph{Physical review B}, 31\penalty0 (8):\penalty0 5262, 1985.

\bibitem[Szlachta et~al.(2014)Szlachta, Bart{\'o}k, and
  Cs{\'a}nyi]{szlachta2014accuracy}
Wojciech~J Szlachta, Albert~P Bart{\'o}k, and G{\'a}bor Cs{\'a}nyi.
\newblock {Accuracy and transferability of Gaussian approximation potential
  models for tungsten}.
\newblock \emph{Physical Review B}, 90\penalty0 (10):\penalty0 104108, 2014.

\bibitem[Zhou and Jones(2011)]{zhou2011effects}
XW~Zhou and RE~Jones.
\newblock {Effects of cutoff functions of Tersoff potentials on molecular
  dynamics simulations of thermal transport}.
\newblock \emph{Modelling and Simulation in Materials Science and Engineering},
  19\penalty0 (2):\penalty0 025004, 2011.

\bibitem[Hinton et~al.(2012)Hinton, Deng, Yu, Dahl, Mohamed, Jaitly, Senior,
  Vanhoucke, Nguyen, Sainath, et~al.]{hinton2012deep}
Geoffrey Hinton, Li~Deng, Dong Yu, George~E Dahl, Abdel-rahman Mohamed, Navdeep
  Jaitly, Andrew Senior, Vincent Vanhoucke, Patrick Nguyen, Tara~N Sainath,
  et~al.
\newblock {Deep neural networks for acoustic modeling in speech recognition:
  The shared views of four research groups}.
\newblock \emph{IEEE Signal processing magazine}, 29\penalty0 (6):\penalty0
  82--97, 2012.

\bibitem[Yegnanarayana(2009)]{yegnanarayana2009artificial}
B~Yegnanarayana.
\newblock \emph{{Artificial neural networks}}.
\newblock PHI Learning Pvt. Ltd., 2009.

\bibitem[LeCun et~al.(2015)LeCun, Bengio, and Hinton]{lecun2015deep}
Yann LeCun, Yoshua Bengio, and Geoffrey Hinton.
\newblock {Deep learning}.
\newblock \emph{Nature}, 521\penalty0 (7553):\penalty0 436, 2015.

\bibitem[Cs{\'a}ji(2001)]{Csaji:Thesis:2001}
Bal{\'a}zs~Csan{\'a}d Cs{\'a}ji.
\newblock {Approximation with artificial neural networks}.
\newblock Master's thesis, Etvs Lornd University, Hungary, 2001.

\bibitem[Bholoa et~al.(2007)Bholoa, Kenny, and Smith]{bholoa2007new}
Ajeevsing Bholoa, Steven~D Kenny, and Roger Smith.
\newblock {A new approach to potential fitting using neural networks}.
\newblock \emph{Nuclear instruments and methods in physics research section B:
  Beam interactions with materials and atoms}, 255\penalty0 (1):\penalty0 1--7,
  2007.

\bibitem[Lorenz et~al.(2004)Lorenz, Gro{\ss}, and
  Scheffler]{lorenz2004representing}
S{\"o}nke Lorenz, Axel Gro{\ss}, and Matthias Scheffler.
\newblock {Representing high-dimensional potential-energy surfaces for
  reactions at surfaces by neural networks}.
\newblock \emph{Chemical Physics Letters}, 395\penalty0 (4-6):\penalty0
  210--215, 2004.

\bibitem[Blank et~al.(1995)Blank, Brown, Calhoun, and Doren]{blank1995neural}
Thomas~B Blank, Steven~D Brown, August~W Calhoun, and Douglas~J Doren.
\newblock {Neural network models of potential energy surfaces}.
\newblock \emph{The Journal of chemical physics}, 103\penalty0 (10):\penalty0
  4129--4137, 1995.

\bibitem[Prudente and Neto(1998)]{prudente1998fitting}
Frederico~V Prudente and JJ~Soares Neto.
\newblock {The fitting of potential energy surfaces using neural networks.
  Application to the study of the photodissociation processes}.
\newblock \emph{Chemical physics letters}, 287\penalty0 (5-6):\penalty0
  585--589, 1998.

\bibitem[Bittencourt et~al.(2004)Bittencourt, Prudente, and
  Vianna]{bittencourt2004fitting}
Ana Carla~P Bittencourt, Frederico~V Prudente, and Jos{\'e} David~M Vianna.
\newblock {The fitting of potential energy and transition moment functions
  using neural networks: transition probabilities in OH (A2$\Sigma$+→
  X2$\Pi$)}.
\newblock \emph{Chemical physics}, 297\penalty0 (1-3):\penalty0 153--161, 2004.

\bibitem[Brown et~al.(1996)Brown, Gibbs, and Clary]{brown1996combining}
David~FR Brown, Mark~N Gibbs, and David~C Clary.
\newblock {Combining ab initio computations, neural networks, and diffusion
  Monte Carlo: An efficient method to treat weakly bound molecules}.
\newblock \emph{The Journal of chemical physics}, 105\penalty0 (17):\penalty0
  7597--7604, 1996.

\bibitem[Bebis and Georgiopoulos(1994)]{bebis1994feed}
George Bebis and Michael Georgiopoulos.
\newblock {Feed-forward neural networks}.
\newblock \emph{IEEE Potentials}, 13\penalty0 (4):\penalty0 27--31, 1994.

\bibitem[Hochreiter(1998)]{hochreiter1998vanishing}
Sepp Hochreiter.
\newblock {The vanishing gradient problem during learning recurrent neural nets
  and problem solutions}.
\newblock \emph{International Journal of Uncertainty, Fuzziness and
  Knowledge-Based Systems}, 6\penalty0 (02):\penalty0 107--116, 1998.

\bibitem[Rumelhart et~al.(1986)Rumelhart, Hinton, and
  Williams]{rumelhart1986learning}
David~E Rumelhart, Geoffrey~E Hinton, and Ronald~J Williams.
\newblock {Learning representations by back-propagating errors}.
\newblock \emph{nature}, 323\penalty0 (6088):\penalty0 533, 1986.

\bibitem[Robbins and Monro(1985)]{robbins1985stochastic}
Herbert Robbins and Sutton Monro.
\newblock {A stochastic approximation method}.
\newblock In \emph{{Herbert Robbins Selected Papers}}, pages 102--109.
  Springer, 1985.

\bibitem[Chollet(2015)]{chollet2015keras}
Fran\c{c}ois Chollet.
\newblock {Keras}.
\newblock \url{https://github.com/fchollet/keras}, 2015.

\bibitem[Abadi et~al.(2016)Abadi, Barham, Chen, Chen, Davis, Dean, Devin,
  Ghemawat, Irving, Isard, et~al.]{abadi2016tensorflow}
Mart{\'i}n Abadi, Paul Barham, Jianmin Chen, Zhifeng Chen, Andy Davis, Jeffrey
  Dean, Matthieu Devin, Sanjay Ghemawat, Geoffrey Irving, Michael Isard, et~al.
\newblock {Tensorflow: a system for large-scale machine learning.}
\newblock In \emph{{OSDI}}, volume~16, pages 265--283, 2016.

\bibitem[Stende(2017)]{Stende:Thesis:2017}
John-Anders Stende.
\newblock {Constructing high-dimensional neural network potentials for
  molecular dynamics}.
\newblock Master's thesis, University of Oslo, Norway, 2017.

\bibitem[Glorot and Bengio(2010)]{glorot2010understanding}
Xavier Glorot and Yoshua Bengio.
\newblock {Understanding the difficulty of training deep feedforward neural
  networks}.
\newblock In \emph{{Proceedings of the thirteenth international conference on
  artificial intelligence and statistics}}, pages 249--256, 2010.

\end{thebibliography}

\end{document}